\documentclass{mn2e}

\usepackage{epsfig}
\title{The Quasar Mass-Luminosity Plane I: A Sub-Eddington Limit for Quasars}
\author[Charles L. Steinhardt and Martin Elvis]
       {Charles L. Steinhardt and Martin Elvis \\
        Harvard-Smithsonian Center for Astrophysics, 60 Garden St, Cambridge, MA 02138}
\date{October 26, 2009}

\begin{document}

\maketitle

\label{firstpage}

\begin{abstract}
We use 62,185 quasars from the Sloan Digital Sky Survey DR5 sample to explore the relationship between black hole mass and luminosity.  Black hole masses were estimated based on the widths of their H$\beta$, Mg{\small II}, and C{\small IV} lines and adjacent continuum luminosities using standard virial mass estimate scaling laws.  We find that, over the range $0.2 < z < 4.0$, the most luminous low-mass quasars are at their Eddington luminosity, but the most luminous high-mass quasars in each redshift bin fall short of their Eddington luminosities, with the shortfall of order ten or more at $0.2 < z < 0.6$.  We examine several potential sources of measurement uncertainty or bias and show that none of them can account for this effect.  We also show the statistical uncertainty in virial mass estimation to have an upper bound of $\sim 0.15$ dex, smaller than the 0.4 dex previously reported.  We also examine the highest-mass quasars in every redshift bin in an effort to learn more about quasars that are about to cease their luminous accretion.  We conclude that the quasar mass-luminosity locus contains a number of new puzzles that must be explained theoretically.
\end{abstract}


\section{Introduction}
\label{sec:intro}

Supermassive black holes (SMBH), with masses between $\sim 10^6 M_\odot$ and $\sim 10^9 M_\odot$, are found at the center of nearly every galaxy where there have been sensitive searches.  While we suspect that the seeds for these SMBH might all have a common origin, their formation mechanism is not well understood.  Many galaxies at redshifts $z \sim 2$ contain quasars, i.e., SMBH in the midst of luminous accretion.  The Soltan argument \cite{Merritt2001} suggests black hole masses are largely accounted for via growth due to luminous accretion.  There are far fewer quasars at low redshift \cite{Schmidt1983,Richards2006}, implying that at some point, SMBH cease their luminous accretion.  The quasar {\em turnoff} mechanism is not well understood (Thacker et al. 2006\nocite{turnoffreview}).  Finally, the black hole mass - stellar velocity ($M-\sigma$) relation \cite{msigma1,msigma2} suggests that SMBH are in some way co-evolving with their host galaxies, but the $M-\sigma$ relation merely describes an end state and is not a theoretical explanation.

As we know of no rapid process by which SMBH can lose mass, the evolutionary tracks for SMBH consist of stages at progressively higher masses.  These stages must, in order, involve (1) a formation mechanism, (2) a period of luminous growth (the `quasar phase'), perhaps along with a period of nonluminous growth (`turnoff'), and (3) a period in which SMBH lie at the centers of galaxies without substantial growth, as we observe them today.  
Much current theoretical research concerns the origin and turnoff phases of quasar evolution, while the `quasar phase' appears to be relatively well understood.  

In this paper we find a new feature of SMBH evolution during their quasar phase.  We examine the evolution of the quasar locus in mass-luminosity space as a function of redshift.  The $M-L$ locus is traditionally shown with all quasars on the same plot (as in Figure \ref{fig:allml}).  The large size of the Sloan Digital Sky Survey (SDSS) DR5 quasar catalogue \cite{SDSSDR5} allows a subdivision into several redshift bins, each containing thousands of quasars.

The Eddington limit produces an absolute upper bound on the luminosity of quasars which is proportional to the black hole mass, $L_{Edd}=1.3\times 10^{46}(M/10^8M_\odot)~{\rm erg~s^{-1}}$ \cite{Shapiro}.  While strictly applicable only for a spherical accretion flow of ionized gas, models for more realistic accretion configurations with rotation are still limited by a luminosity of this order
(except under special circumstances, e.g. Begelman 2002)\nocite{Begelman}.  Using a smaller data set ($N = 733$) than SDSS, Kollmeier et al. (2006)\nocite{Kollmeier2005} appeared to confirm the applicability of $L_{Edd}$ to quasars, using an $M-L$ locus (similar to Figure \ref{fig:allml}) to show that the most luminous quasars reach but do not exceed $L_{Ed}$.  For the lowest black-hole masses at every redshift, we confirm the conclusion of \cite{Kollmeier2005}, but we show that the quasars with the highest SMBH mass at every redshift fall well short of $L_{Edd}$. 

In {\S~\ref{sec:methods}}, we review the methods used to estimate masses and bolometric luminosities.  While we have added nothing original to this methodology, the remainder of our results are entirely dependent upon its accuracy.  In {\S~\ref{sec:lm}}, we subdivide the SDSS DR5 quasar catalogue by redshift and show that the quasar mass-luminosity distribution does not match what we should expect given our current theoretical understanding of quasar accretion.  In particular, we show that instead a sub-Eddington boundary (SEB) is present in each redshift bin.  We also use our limited statistics to make a first estimate for how the SEB evolves with redshift.  In {\S~\ref{sec:tests}}, we consider several alternative explanations for the discrepancies between the observed quasar locus and the Eddington luminosity, focusing on potential sources of measurement uncertainty or bias.  In {\S~\ref{sec:discussion}}, we comment on the potential implications of our new results.

\section{Virial Mass Estimation}
\label{sec:methods}

The remainder of this work relies upon an ability to accurately estimate the bolometric luminosities and central black hole masses of quasars at cosmological redshifts.  The primary results in this work are drawn from the Shen et al. (2008)\nocite{Shen2008} virial mass catalogue for SDSS DR5 \cite{SDSSDR5} quasars.  The luminosity determination is fairly straightforward and uses the relatively settled techniques discussed in Richards et al. (2006)\nocite{Richards2006}.  The mass determination, on the other hand, uses relatively new techniques, and has a larger uncertainty.  Therefore, we review here the basic assumptions in virial mass estimation.  Potential sources of error or bias in the Shen et al. virial masses are discussed in more detail in {\S~\ref{sec:tests}}

The determination of black hole masses from spectral emission lines relies upon two basic assumptions: (1) that the orbital velocity of gas in the broad-line region (BLR) is dominated by the virial velocity due to the central black hole and (2) that there is a scaling relationship between luminosity and radius.  As a result of (1), we can calculate the mass $M_{BH}$ of the black hole from emission lines in the BLR as 
\begin{equation}
\label{eq:virial}
M_{BH} = \frac{R_{\textrm{BLR}}v_{\textrm{BLR}}^2}{G},
\end{equation}
where $R_{\textrm{BLR}}$ is the radius and $v_{\textrm{BLR}}$ the velocity of gas emitting the BLR spectral lines.  
Marconi et al. (2009)\nocite{Marconi2009} suggest corrections to the virial approximation for radiation pressure might be needed at high luminosity, particularly when using C{\small IV} lines to determine $v_{\textrm{BLR}}$.
This first scaling relationship is based upon black hole masses determined using reverberation mapping, which uses the time delay between variability in the continuum and emission lines to determine $R_{\textrm{BLR}}$ (cf. Peterson \& Horne 2004)\nocite{Peterson2004}.

Virial mass estimates also require assumption (2), an empirical scaling relationship very close to the $L \alpha R^2$ that we would expect for a black body \cite{Bentz2009}, although thermal processes in the accretion disc are likely to be substantially more complex.  This second assumption transforms (\ref{eq:virial}) into a scaling relation of the form
\begin{eqnarray}
\label{vmeform}
\log (M/M_\odot) & = & A \\ & + & \log
\left[\left(\frac{\textrm{FWHM}(\textrm{H}\beta)}{1000\textrm{
km/s}}\right)^2\left(\frac{\lambda L_\lambda (B~{\rm
\AA})}{10^{44}\textrm{ erg/s}}\right)^{C}\right] \nonumber,
\end{eqnarray}
where FWHM is the Full Width at Half-Maximum of the corresponding line profile, $L_{\lambda}$ is the luminosity per unit wavelength at rest-frame wavelength $\lambda = B$, and $A, B,$ and $C$ are constants.  Vestergaard \& Peterson (2006)\nocite{Vestergaard2006} determined these constants in black hole mass scaling relations based on H$\beta$ and C{\small IV} emission lines (VP06), while McLure \& Jarvis (2002) and McLure \& Dunlop (2004) developed a mass relation (MD04) by scaling Mg{\small II}-based estimates against H$\beta$-based estimates.  The scaling relations used in this work are summarized in Table \ref{table:vmesummary}.  Multiple scaling relations are required because SDSS spectra only cover the range 3900--9100 \AA, so that none of these three BLR emission lines are accessible over the entire redshift range of the SDSS catalogue.

The additive constants $A$ are determined by calibrating virial mass scaling relationships against other methods for black hole mass estimation as part of a `black hole mass ladder'\cite{Peterson2004}.  Direct estimates based upon local stellar and gas kinematics (cf. Ferrarese \& Ford 2005\nocite{Ferrarese2005} provide our best estimates for nearby black holes.  Reverberation mapping is calibrated against these estimates, and the virial mass scaling relations are in turn calibrated against reverberation mapping.  These calibrations use very few quasars compared to the 62,185 in the Shen et al. (2008) sample: only 28 quasars are used to calibrate the H$\beta$ and 27 to calibrate the C{\small IV} scaling relations\cite{Vestergaard2006}. 
 
\begin{table}
\caption{Summary of the virial mass estimates used}
\begin{tabular}{|c|c|c|c|c|c|}
\hline 
Sample & $A$ & $B$ & $C$ & Redshift & Source\\
\hline 
H$\beta$ mass & 6.91 & 5100 & 0.50 & 0--0.872 & VP06 \\
Mg{\small II} mass & 6.51 & 3000 & 0.50 & 0.393 -- 2.252 & MD04 \\
C{\small IV} mass & 6.66 & 1350 & 0.53 & 1.518--4.875 & VP06\\
\hline  
\end{tabular}
\label{table:vmesummary}
\end{table}

There is a small overlap in redshift between the H$\beta$ mass and Mg{\small II} mass samples as well as one between the Mg{\small II} mass and C{\small IV} mass samples (Table \ref{table:vmesummary}).  Shen et al. (2008) show that the agreement between mass estimates for the same object taken using the H$\beta$ and Mg{\small II} relations (0.22 dex dispersion), is better than between estimates using the Mg{\small II} and C{\small IV} relations (0.34 dex, and correlated with the C{\small IV}-Mg{\small II} blueshift).  The C{\small IV} mass estimates may be less accurate and we therefore focus here almost exclusively on masses obtained using the other two scaling relations.  We consider the implications of this disagreement further in \S~{\ref{sec:tests}}.

\section{The Mass-Luminosity Relation}
\label{sec:lm}
Black hole masses for 62,185 of the 77,429 SDSS DR5 quasars were determined by Shen et al. (2008) using the scaling relations summarized in Table \ref{table:vmesummary}: 10,605 in the H$\beta$-mass sample, 42,035 in the Mg{\small II}-mass sample, and 14,565 in the C{\small IV}-mass sample.  The catalogue includes 3505 quasars with both H$\beta$ and Mg{\small II} masses and 3427 objects with both Mg{\small II} and C{\small IV} masses.  Because the C{\small IV}-based mass estimates may be less accurate, our main results are derived from 49,135 DR5 quasars in the H$\beta$ and Mg{\small II} mass samples.

\begin{figure}
\leavevmode
      \epsfxsize=3in\epsfbox{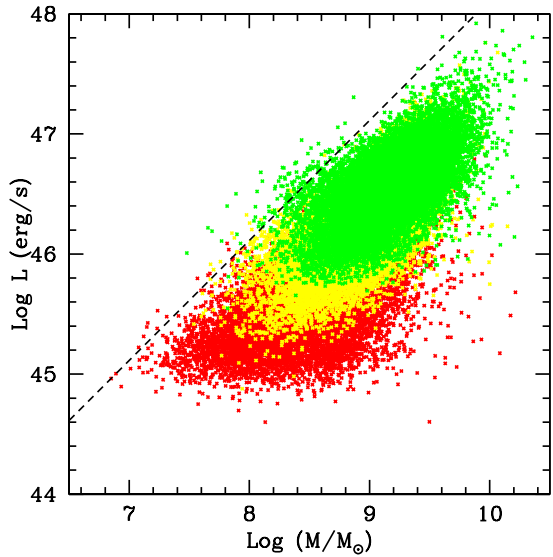}
  \caption{The quasar locus in the mass-luminosity plane for all quasars from $0.2 < z < 2.0$, using virial masses estimated by Shen et al. (2008) with H$\beta$ and Mg{\small II} lines and bolometric luminosities using the techniques of Richards et al. (2006)  The dashed line is drawn at the Eddington luminosity as a function of mass.  The colour indicates whether the quasar is at $0.2 < z < 0.8$ (red), $0.8 < z < 1.4$ (yellow), or $1.4 < z < 2.0$ (green).}
\label{fig:allml}
\end{figure}

Figure \ref{fig:allml} displays virial mass estimates and bolometric luminosities for all quasars in the H$\beta$ and Mg{\small II} mass samples.
The most striking feature is that the quasar locus seems bounded by $L_{Edd}$ (dashed line), as already shown by Kollmeier et al. (2006)\nocite{Kollmeier2005}.  However, while the $L_{Edd}$ bound is tight at most masses, we also note a slight departure from the $L_{Edd}$ bound for $M > 10^9 M_\odot$.  It is also apparent that the locus of quasars at $1.4 < z < 2.0$ (green) is different than the locus at $0.2 < z < 0.8$ (red).  A proper investigation of this possible `sub-Eddington boundary' (hereafter SEB) therefore requires a subdivision of Figure \ref{fig:allml} into redshift bins. 

\subsection{Quasar mass evolution}

\begin{figure}
\leavevmode
\includegraphics[width=3in,height=4in]{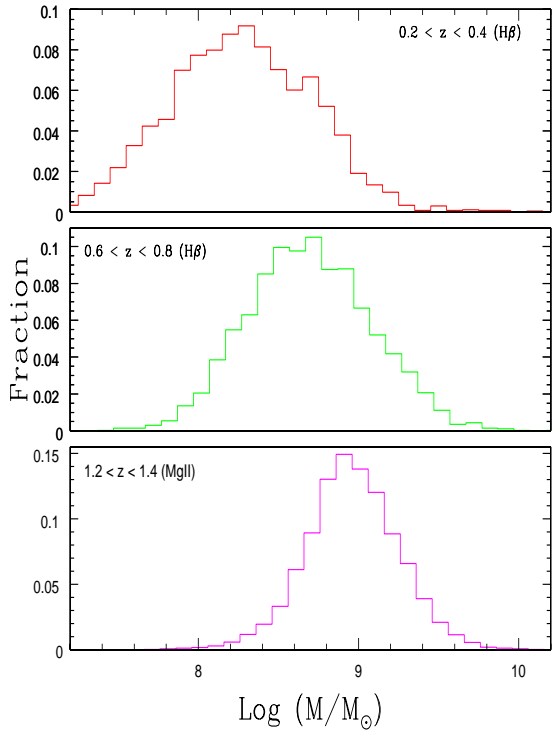}
  \caption{The mass distribution for three different redshift bins within
  our sample.  The red ($0.2 < z < 0.4$) and green ($0.6 < z < 0.8$) lines
  are from the H$\beta$ mass sample and the purple ($1.2 < z < 1.4$) line is from the Mg{\small II} mass sample.  These mass distributions are not completeness-corrected, and the low-mass tails of these distributions are likely lowered by SDSS magnitude selection.  However, high-mass quasars can be detected even at low $L_{Edd}$, so a lack of high-mass quasars at low redshifts cannot be ascribed to selection.}
\label{fig:masshist}
\end{figure}

We have therefore divided the H$\beta$ mass and Mg{\small II} mass samples into 10 redshift bins of size $0.2$ in $z$.  Table {\ref{table:bins}} contains summary statistics on the objects in each bin.  Within each bin, there is a distribution of black hole masses spanning $\sim 2$ dex, as shown in Figure \ref{fig:masshist}.   
\begin{table}
\caption{Summary statistics on quasars in the 10 redshift and emission line bins}
\begin{tabular}{|c|c|c|c|c|c|c|c|}
\hline 
ID & $z$ & $N$ & $<\log L>$ & $\sigma_L$ & $<\log M/M_\odot>$ & $\sigma_M$ \\
\hline 
H$\beta$ & & & & & & \\
1 & 0.2-0.4 & 2690 & 45.25 & 0.20 & 8.27 & 0.44 \\
2 & 0.4-0.6 & 4250 & 45.54 & 0.25 & 8.44 & 0.42 \\
3 & 0.6-0.8 & 3665 & 45.89 & 0.25 & 8.69 & 0.39 \\
Mg{\small II} & & & & & & \\
4 & 0.6-0.8 & 4727 & 45.80 & 0.29 & 8.59 & 0.32 \\
5 & 0.8-1.0 & 5197 & 46.02 & 0.30 & 8.76 & 0.31 \\
6 & 1.0-1.2 & 6054 & 46.21 & 0.26 & 8.89 & 0.29 \\
7 & 1.2-1.4 & 7005 & 46.32 & 0.27 & 8.96 & 0.29 \\
8 & 1.4-1.6 & 7513 & 46.43 & 0.27 & 9.07 & 0.28 \\
9 & 1.6-1.8 & 6639 & 46.57 & 0.24 & 9.18 & 0.29 \\
10 & 1.8-2.0 & 4900 & 46.71 & 0.22 & 9.29 & 0.30 \\
%11 & 1.8-2.0 & 4627 & 46.60 & 9.20 & C{\small IV} \\
%12 & 2.0-3.0 & 7079 & 46.79 & 9.30 & C{\small IV} \\
%13 & 3.0-4.1 & 2859 & 46.98 & 9.38 & C{\small IV} \\
\hline  
\end{tabular}
\label{table:bins}
\end{table}

It has been known for a long time that quasars are `downsizing', or that the brightest quasars at higher-redshift are more intrinsically luminous than the brightest quasars at lower-redshift \cite{Schmidt1968}.  Similarly, quasars above the peak in each of the higher-redshift mass distributions in Figure \ref{fig:masshist} lie at masses with substantially smaller populations in the quasar mass distributions at lower redshift.  This shows that many higher-mass quasars turn off by lower redshift (and disappear from the sample) rather than become less luminous (but remain in the sample at lower luminosity).  We discuss quasar turnoff in more detail in Paper II\nocite{Steinhardt2009b}.  While the low-mass quasar distribution also varies with redshift, the SDSS detection limit is at a fixed magnitude.  Therefore, lower-mass and more distant quasars must be closer to their Eddington luminosity in order to be bright enough to be include in the SDSS catalogue.  As a result, the low-mass tails of these mass distributions may be skewed by SDSS selection.  However, high-mass quasars can be detected even at low $L_{Edd}$, and a lack of high-mass quasars at low redshift cannot be ascribed to selection.  

\subsection{The Mass-Luminosity Plane at $0.2 < z < 0.4$}
\label{ssec:defab}

The brightest quasars are more intrinsically luminous at higher redshift, and similarly Figure \ref{fig:masshist} demonstrates that the biggest central black holes are more massive at higher redshift.  Further, Figure \ref{fig:allml} demonstrates that the most luminous quasars at every mass are near $L_{Edd}$.  It is therefore natural to believe that quasar luminosity downsizing and quasar mass downsizing are simultaneous, such that the most massive and most luminous quasars decline with equal speed in both mass and luminosity towards lower redshift, remaining near $L_{Edd}$ at every redshift.  

\begin{figure}
  \epsfxsize=3in\epsfbox{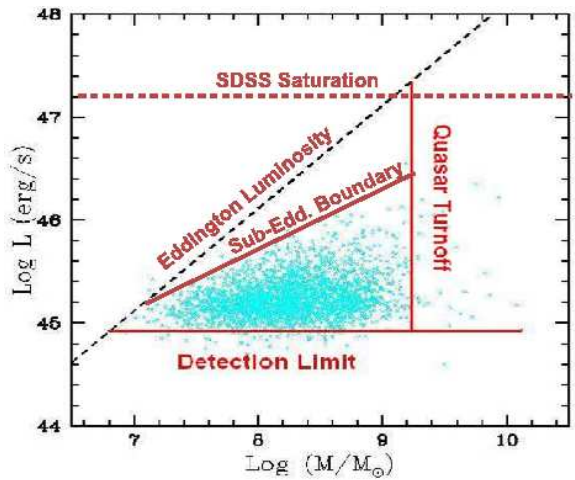}
\caption{The SDSS quasar locus of the $H\beta$ mass sample in the $M-L$ plane at redshift $0.2 < z < 0.4$.  The locus should be bounded by SDSS {\bf detection limits}, $L_{Edd}$ (dashed line), and on the high-mass end by an unknown mechanism responsible for {\bf quasar turnoff}.  In practice, there appears to be an additional sub-Eddington boundary with slope below that of $L_{Edd}$.  The bright-object {\bf SDSS saturation} limit does not intersect the quasar locus.}
\label{fig:hb1}
\end{figure}
In Figure \ref{fig:hb1}, we present the quasar locus at $0.2 < z < 0.4$ in the $M-L$ plane.  In some papers this plane is plotted with luminosity on the abscissa.  We prefer to put mass on the abscissa as mass is a less variable property of the object.  The origin of the boundaries of this locus are mostly understood:  The SDSS DR5 selection has magnitude limits due to detector sensitivity at $i \sim 22$ (which we have labeled as {\bf Detection limit} in Figure \ref{fig:hb1}) and {\bf SDSS saturation} at $i < 16$ (which does not bound any of our quasar loci).  
There is also a high-mass limit.  As larger SMBH do exist at higher redshift, we have labeled this limit as the {\bf Quasar turnoff}.  This limit is discussed in detail in Paper II \cite{Steinhardt2009b}.  The dashed line is drawn at $L_{Edd}(M)$.  As in Figure \ref{fig:allml}, there are no quasars statistically exceeding $L_{Edd}$.  

More strikingly, there is a tighter bound than $L_{Edd}$ on the maximum luminosity at masses $M > 10^{7.5} M_\odot$.  We term this the {\em Sub-Eddington Boundary} (SEB; the red line in Figure \ref{fig:hb1}).  This SEB is much more prominent than in the entire sample (Figure \ref{fig:allml}), implying a redshift evolution of the SEB.  To determine the shape of the SEB, we consider the quasar luminosity distribution as a function of mass.  
\begin{figure}
  \epsfxsize=3in\epsfbox{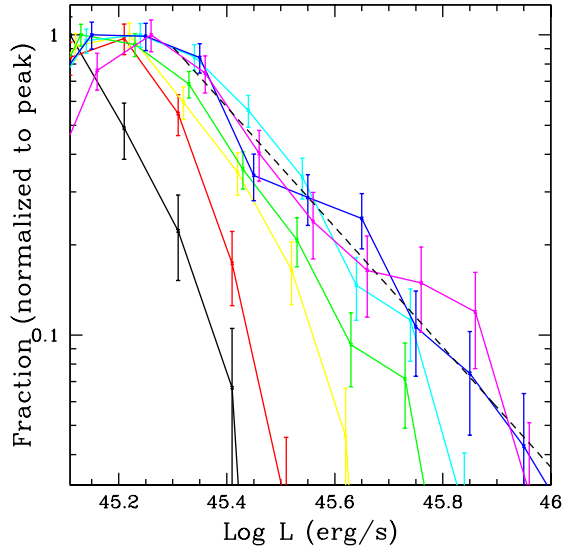}
\caption{The quasar number density for the quasars shown in Figure \ref{fig:hb1} at $0.2 < z < 0.4$ as a function of luminosity in seven different mass slices, each of width 0.25 dex in $\log M/M_\odot$: black (7.25-7.5), red (7.5-7.75), yellow (7.75-8.0), green (8.0-8.25), cyan (8.25-8.5), blue (8.5-8.75), and purple (8.75-9.0).  The dashed line is drawn proportional to $L^{-2}$ and is normalized to the purple (8.75-9.0) curve.}
\label{fig:mvslslice}
\end{figure}
Figure \ref{fig:mvslslice} shows the quasar number density as a function of luminosity in different mass bins including Poisson errors.  The distributions in each mass bin have a peak number density with a decline at high luminosity (see \S~\ref{sec:tests}) described in Table \ref{table:lumdropoff}.  
\begin{table}
\caption{Best-fitting exponential decays $N \propto L^{-k}$ for the quasar luminosity function in different mass bins at $0.2 < z < 0.4$}
\begin{center}
\begin{tabular}{|c|c|c|}
\hline 
$\log M/M_\odot$ & slope $k$ & $\chi^2$/DOF \\
\hline 
7.25-7.5 & $3.69 \pm 0.33$ & 0.41 \\
7.5-7.75 & $4.69 \pm 0.59$ & 3.23 \\
7.75-8.0 & $2.96 \pm 0.35$ & 1.80 \\
8.0-8.25 & $2.01 \pm 0.22$ & 4.77 \\
8.25-8.5 & $2.25 \pm 0.22$ & 3.74 \\
8.5-8.75 & $1.79 \pm 0.15$ & 3.62 \\
8.75-9.0 & $1.96 \pm 0.17$ & 1.01 \\
\hline  
\end{tabular}
\end{center}
\label{table:lumdropoff}
\end{table}  
At high mass, where a wide range of luminosity is visible above the peak number density and the maximum luminosity is sub-Eddington, the best-fitting exponential decay is $\sim L^{-2}$.  At low mass, where SDSS detection limits the sample to a narrow range of luminosity and the most luminous objects lie near $L_{Edd}$, the quasar number density decline may be steeper with increasing luminosity than at high mass.  A majority of the best-fitting declines have $\chi^2$/DOF $ > 3$, suggesting that the falloff may not be purely exponential.  Since the quasars at lowest mass only allow a fit from 4-5 points, it is also possible that these declines are all $\sim L^{-2}$ for much of their luminosity range but steeper near the SEB.

The peaks in each mass bin do not lie at the SDSS low-magnitude cutoff, but rather are true peaks in the luminosity number density.  The quasar luminosity function declines as a power law in luminosity $\sim L^{-2}$ (cf. Richards et al. 2006b\nocite{Richards2006b}, Amarie et al. 2009\nocite{Amarie2009}) at luminosities above the peak number density.  For comparison, a decline proportional to $L^{-2}$ is shown in Figure \ref{fig:mvslslice}.  Since the $\sim 0.4$ dex mass uncertainty is large compared to the $0.25$ dex mass bins, a majority of quasars should randomly lie in the wrong mass bin and $\sim 1/3$ of quasars would not even lie in a bin adjacent to their correct mass bin.  Detailed fit parameters for the decline are likely not credible and have a high $\chi^2$/DOF as in Table \ref{table:lumdropoff}, but a power-law decline $\sim L^{-2}$ is consistent with the data in each of the higher mass bins where quasars take on the widest range of luminosity.

\begin{figure}
  \epsfxsize=3in\epsfbox{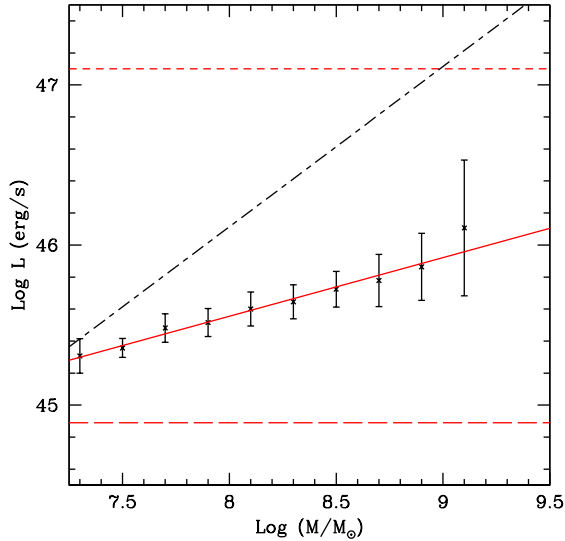}
\caption{The SEB as approximated by the 95th percentile luminosity above peak number density at $0.2 < z < 0.4$.  The best-fitting linear approximation has slope is $\alpha = 0.37 \pm 0.02$, well below the $\alpha=1$ slope of $L_{Edd}$ (white, alternating dashes).  The indicated uncertainties are derived via bootstrapping and the fit has a very low $\chi^2$/DOF of 0.06.  Note that the SEB at every mass lies well away from the SDSS detection and saturation limits (long and short dashed lines, respectively).}
\label{fig:bestfit}
\end{figure}

We define the luminosity boundary $L_{cutoff}(M)$ robustly in each mass bin as the 95th percentile quasar luminosity above the peak.  In each mass bin, the uncertainty can be estimated using bootstrapping (repeatedly choosing a set of $N$ quasars randomly with replacement from the $N$ quasars in each mass bin), finding the standard deviation of the resulting 95th percentile luminosities.  
Figure {\ref{fig:bestfit}}, shows $L_{cutoff}(M)$ for the $0.2 < z < 0.4$ redshift bin.  The best straight-line fit for $L_{cutoff}$ has slope $\alpha = 0.37 \pm 0.02$, well below $\alpha_{Edd} = 1$ (Figure {\ref{fig:hb1}}).  An SEB is strongly required.  The low $\chi^2$/DOF of 0.06 for this linear fit is likely evidence of correlated uncertainties, perhaps due to the large mass uncertainty placing individual objects in incorrect bins.  As shown in \S~\ref{ssec:twocomp}, $\chi^2$/DOF for this redshift bin is atypically low.

The highest-mass bins lie further from our best-fitting lines but also have larger uncertainties due to a lower quasar number density, while quasars with extremal mass estimates are most likely to have been placed in the wrong bin.  It is possible that the SEB would be well-fit by two linear components given lower-uncertainty measurements of $M$ and $L$.  We consider whether such measurement uncertainties might be responsible for the $\alpha < 1$ slope of the SEB in \S~{\ref{sec:tests}}.  

\subsection{Evolution of the sub-Eddington boundary}
\label{ssec:twocomp}

\begin{figure}
  \epsfxsize=3in\epsfbox{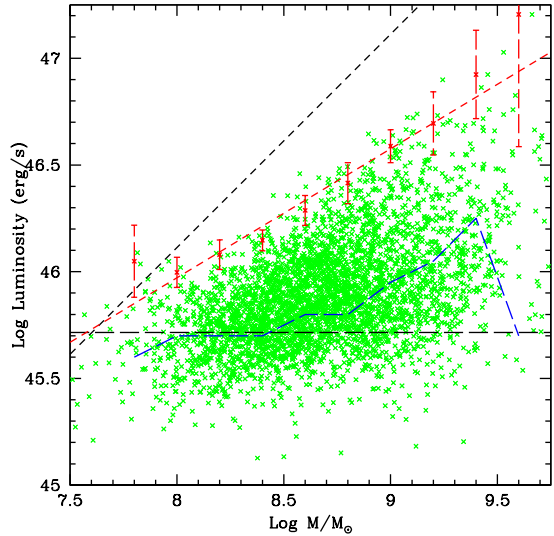}
\caption{The quasar distribution in the mass-luminosity plane at $0.6 < z < 0.8$ as measured using H$\beta$ masses.  The sub-Eddington boundary (red, dashed) is fit from 95th percentile points (red) above the peak number density (blue, dashed) and is different than that at $0.2 < z < 0.4$ (Figure \ref{fig:hb1}) but is still present.  The black dashed line is drawn at a bolometric luminosity approximately corresponding to $i = 19.2$ for a typical quasar SED at this redshift.}
\label{fig:hb3}
\end{figure}

\begin{figure}
  \epsfxsize=3in\epsfbox{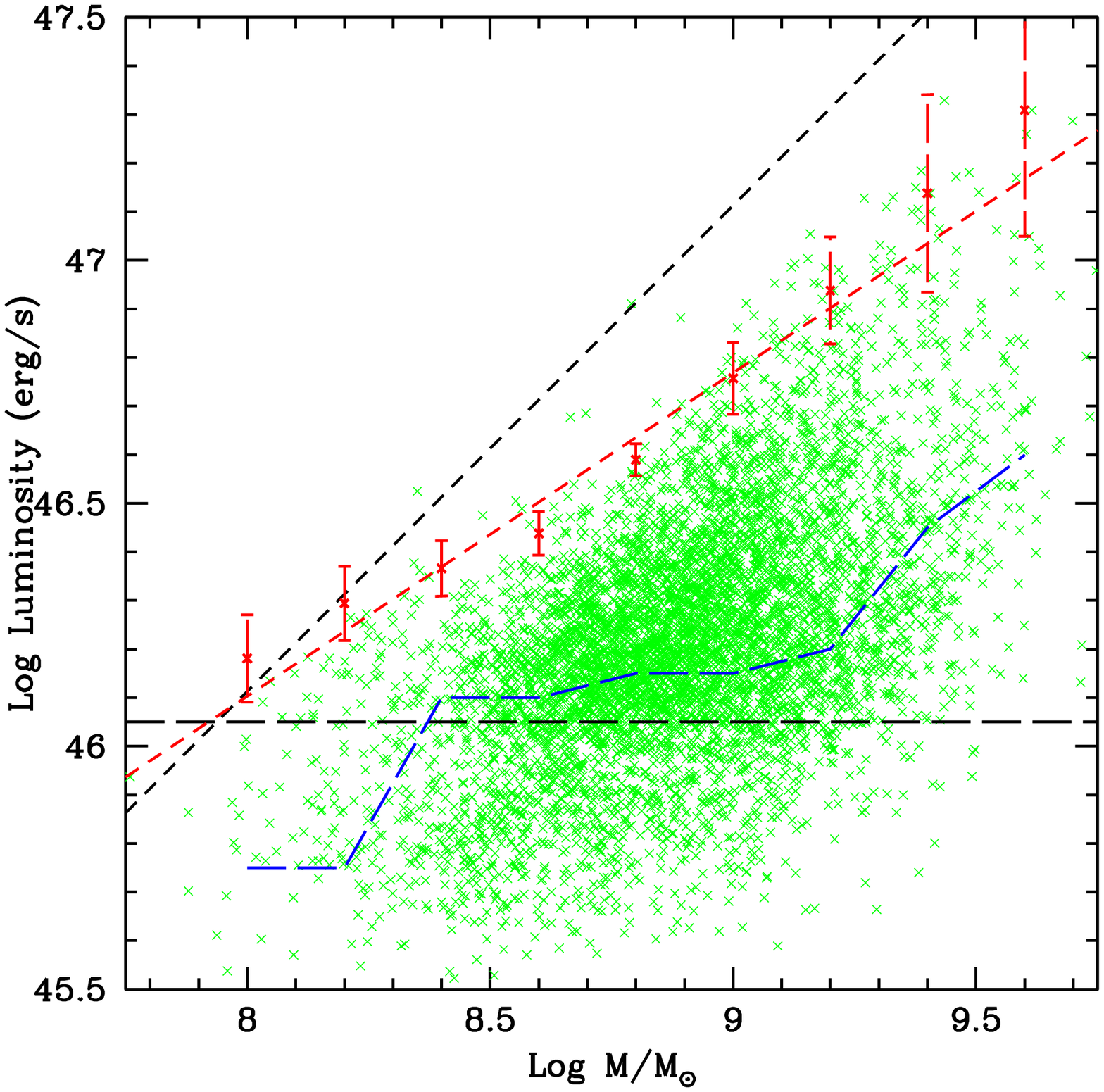}
\caption{The quasar distribution in the mass-luminosity plane at $1.0 < z < 1.2$ as measured using Mg{\small II} masses.  The sub-Eddington boundary (red, dashed) is fit from 95th percentile points (red) above the peak number density (blue, dashed) and has a slope much closer to the Eddington limit than it does at $0.2 < z < 0.4$ (Figure \ref{fig:hb1}) but is still not parallel.  The quasar population approaches $L_{Edd}$ at lower mass but does not at higher mass.  The black dashed line is drawn at a bolometric luminosity approximately corresponding to $i = 19.2$ for a typical quasar SED at this redshift.}
\label{fig:mg2}
\end{figure}

\begin{figure*}
  \epsfxsize=7in\epsfbox{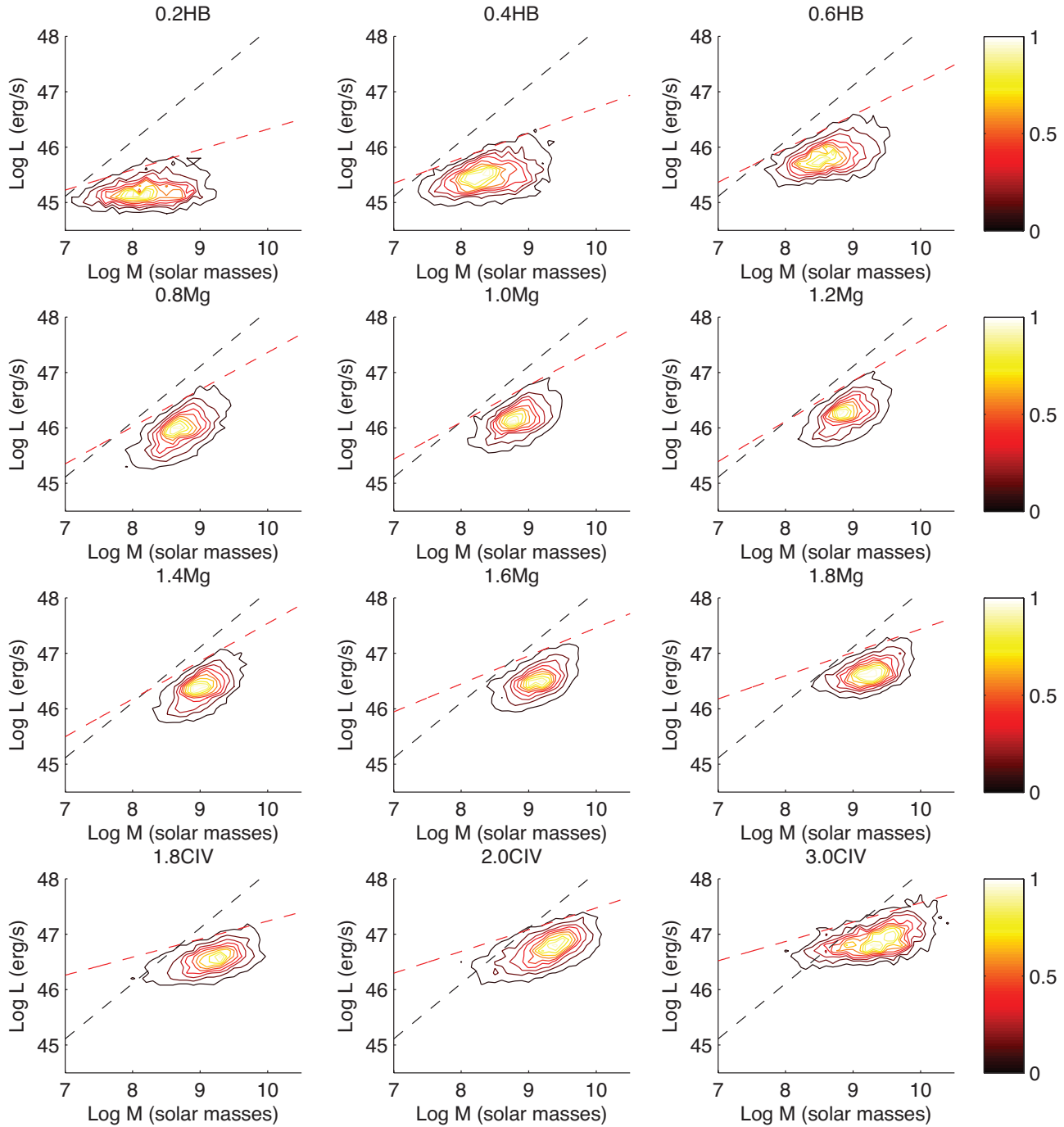}
\caption{A contour plot of the mass-luminosity distribution in 12 different redshift ranges with the best-fitting SEB (red, dashed). In each redshift bin, the quasar number density has been normalized to the peak number density.  The top three panels are from our low-redshift sample with H$\beta$-based mass estimates, the middle six panels are from our medium-redshift sample with Mg{\small II}-based mass estimates, and the bottom three panels are from our high-redshift sample and use C{\small IV}-based mass estimates.  The three redshift intervals with C{\small IV}-based masses suffer from suspected flaws in mass estimation (Shen et al. 2008); however, they are included to suggest, with the best available data, that the SEB continues to higher redshifts.}
\label{fig:allzcontour}
\end{figure*}

Figure \ref{fig:allzcontour} shows quasar loci in the $L-M$ plane for each of the 12 redshift bins as contour plots.  An SEB is detected in each panel, although the location and slope appear to evolve with redshift.  An offset at which quasars fall short of $L_{Edd}$ when using Mg{\small II} masses has been previously reported \cite{Kollmeier2005,Shen2008,Gavignaud2008,Trump2009}.  The effect found here is different as the offset is mass dependent and changes slope with redshift (Figures \ref{fig:hb3}, \ref{fig:mg2}): the maximum Eddington ratio at higher masses is below that at lower masses.

While the three C{\small IV} mass samples have known flaws, they are included in Figure \ref{fig:allzcontour} in order to demonstrate that the best available evidence suggests that the SEB continues to exist at redshifts higher than the $z \sim 2.0$ limitations of SDSS spectra-based Mg{\small II} mass estimation.

In \S~{\ref{ssec:defab}}, we showed that the $0.2 < z < 0.4$ SEB appears to be well-fit by a linear $L = \alpha M + L_0$, as in Figure \ref{fig:hb1}.  We therefore parametrize the SEB in the same manner in each redshift bin.  In Table \ref{table:fitparams}, we show the results of these fits for $0.2 < z < 2.0$. 
\begin{table}
\caption{Fit parameters at different redshift for the maximum luminosity $L = \alpha M + L_0$.  This is a two-parameter fit to typically 8-10 total points.  Both the H$\beta$ and Mg{\small II} views at $0.6 < z < 0.8$ are included.}
\begin{tabular}{|c|c|c|c|c|c|}
\hline 
Redshift & $\alpha$ & $\sigma_\alpha$ & $L_0$ & $\sigma_{L_0}$ & $\chi^2$/DOF \\
\hline 
H$\beta$ \\
0.2-0.4 & 0.37 & 0.02 & 42.63 & 0.18 & 0.06 \\
0.4-0.6 & 0.45 & 0.03 & 42.13 & 0.22 & 0.28 \\
0.6-0.8 & 0.60 & 0.06 & 41.07 & 0.49 & 0.56 \\
Mg{\small II} \\
0.6-0.8 & 0.61 & 0.10 & 41.03 & 0.83 & 1.11 \\
0.8-1.0 & 0.67 & 0.09 & 40.60 & 0.79 & 0.90 \\
1.0-1.2 & 0.67 & 0.05 & 40.71 & 0.48 & 0.85 \\
1.2-1.4 & 0.73 & 0.05 & 40.24 & 0.48 & 1.13 \\
1.4-1.6 & 0.68 & 0.08 & 40.66 & 0.72 & 1.10 \\
1.6-1.8 & 0.50 & 0.10 & 42.35 & 0.89 & 1.64 \\
1.8-2.0 & 0.42 & 0.06 & 43.20 & 0.51 & 1.06 \\
\hline  
\end{tabular}
\label{table:fitparams}
\end{table}
At every redshift, the SEB fit parameters shown in Table \ref{table:fitparams} show a slope at least $3.7\sigma$ below that of the Eddington luminosity.   

These best-fitting parameters depend upon the peak number density.  At most combinations of mass and redshift, the peak number density lies at a luminosity higher than many objects in the SDSS catalog.  However, while the SDSS catalog is nearly complete for objects brigher than $i = 19.2$, fainter objects are only included if they are ``serendipitously'' selected as a ROSAT source, FIRST source, etc.  In Paper II, we compare this serendipitous sample more closely with the remainder of the SDSS catalog and consider their views of the low-luminosity end of the mass-luminosity quasar distribution.  While $i = 19.2$ does not translate directly to a bolometric luminosity because the quasar spectal energy distribution can vary, a bolometric luminosity typical of quasars in the SDSS catalog with $i = 19.2$ is indicated in Figures \ref{fig:hb3} and \ref{fig:mg2}.  It is possible that peak luminosity has been overestimated because of incompleteness below $i = 19.2$.  The entire luminosity distribution at fixed mass and redshift typically spans between 0.8 and 1.5 dex (the implications of this are discussed further in Paper II).  In Table \ref{table:peakshift}, we consider the possibility that the peaks are overestimated due to selection and show the best-fitting SEB as defined using a peak 0.2 dex lower in luminosity at every mass and redshift.  Because of the sharp decline in number density near the SEB, an 0.2 dex shift in the peak corresponds to a smaller shift in the 95th percentile object used to estimate the SEB.  We also consider shifting only peaks at the lowest 1.0 dex of mass at each redshift in an attempt to bias the SEB determination as far as possible towards a slope of 1.

\begin{table}
\caption{Fit parameters at different redshift for the maximum luminosity $L = \alpha M + L_0$ using the 95th percentile objects above or within 0.2 dex of the peak number density at each mass and redshift.  Best-fitting parameters using the SDSS peak at high mass but this shifted peak at low mass are also considered in an attempt to bias the SEB determination as far as possible towards a slope of 1.}
\begin{tabular}{|c|c|c|c|c|c|}
\hline 
Redshift & Original Slope & Slope, 0.2 below Peak & Slope, biased peaks \\
\hline 
H$\beta$ \\
0.2-0.4 & $0.37 \pm 0.02$ & $0.34 \pm 0.03$ & $0.37 \pm 0.02$ \\
0.4-0.6 & $0.45 \pm 0.03$ & $0.48 \pm 0.02$ & $0.49 \pm 0.02$ \\
0.6-0.8 & $0.60 \pm 0.06$ & $0.61 \pm 0.06$ & $0.64 \pm 0.06$ \\
Mg{\small II} \\
0.6-0.8 & $0.61 \pm 0.10$ & $0.60 \pm 0.08$ & $0.64 \pm 0.10$ \\
0.8-1.0 & $0.67 \pm 0.09$ & $0.65 \pm 0.06$ & $0.69 \pm 0.07$ \\
1.0-1.2 & $0.67 \pm 0.05$ & $0.66 \pm 0.06$ & $0.68 \pm 0.06$ \\
1.2-1.4 & $0.73 \pm 0.05$ & $0.72 \pm 0.05$ & $0.74 \pm 0.06$ \\
1.4-1.6 & $0.68 \pm 0.08$ & $0.67 \pm 0.08$ & $0.70 \pm 0.08$ \\
1.6-1.8 & $0.50 \pm 0.10$ & $0.51 \pm 0.09$ & $0.54 \pm 0.10$ \\
1.8-2.0 & $0.42 \pm 0.06$ & $0.42 \pm 0.06$ & $0.45 \pm 0.06$ \\
\hline  
\end{tabular}
\label{table:peakshift}
\end{table}
\begin{figure}
  \epsfxsize=3in\epsfbox{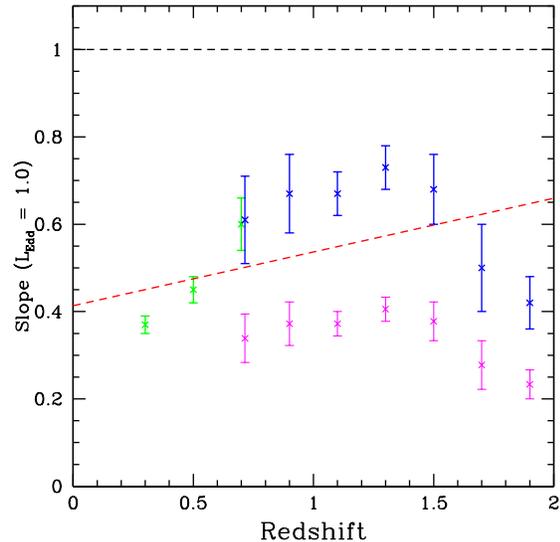}
\caption{Evolution of the best-fitting SEB slope displayed in Table \ref{table:fitparams}.  The green measurements are from the H$\beta$ mass sample, while blue points are from the Mg{\small II} mass sample.  The best-fitting linear evolution for the SEB slope is a poor fit, with $\chi^2$/DOF of 7.84.  The magenta points use possible corrections to Mg{\small II} masses discussed in \S~\ref{subsec:vme}.}
\label{fig:fitevolution}
\end{figure} 
In Figure \ref{fig:fitevolution}, we show the redshift evolution of the SEB slope $\alpha$,.   All of the $\alpha$ lie within a narrow range of standard deviation 0.12.  The best linear fit to the slope as a function of redshift is $\alpha(z) = (0.12 \pm 0.08)z + (0.41 \pm 0.09)$, as in Figure \ref{fig:fitevolution}, but is a poor fit, with $\chi^2$/DOF of 7.84.  This analysis cannot exclude the possibility that the SEB takes on a slope independent of redshift.

\section{Is the Sub-Eddington Boundary Due to Measurement Error?}
\label{sec:tests}

We can divide the explanations for the SEB into four possibilities:
\begin{enumerate}
\item{SDSS selection excluding high-luminosity, high-mass quasars from the DR3 and DR5 catalogues.}
\item{Measurement errors resulting in either an underestimated bolometric luminosity or incorrect fit parameters for the spectral lines used to estimate $M$.}
\item{Incorrect virial mass scaling relations for higher-mass quasars.}
\item{Physical effects limiting luminous quasar accretion more strongly than the Eddington limit.}
\end{enumerate} 
We consider the first three possibilities in this section, with a summary of the potential explanations considered in Table \ref{table:tests}.  We briefly consider the implications of these boundaries being physical in \S~{\ref{sec:discussion}}. A full consideration of possible physical causes for the SEB is beyond the scope of this paper. 

\subsection{Statistical significance}

We first consider whether the SEB is statistically significant, or whether the $L/L_{Edd}$ distribution is consistent with being identical in different mass bins.  
\begin{figure}
  \epsfxsize=3in\epsfbox{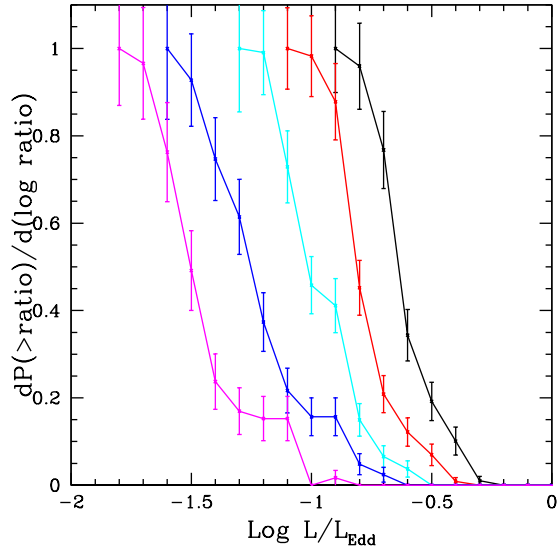}
\caption{The distribution of Eddington ratios in five mass bins of width 0.25 dex in $\log M/M_\odot$ at $0.2 < z < 0.4$ from Figure \ref{fig:mvslslice}: black (7.75-8.0), red (8.0-8.25), cyan (8.25-8.5), blue (8.5-8.75), and purple (8.75-9.0).  The low-Eddington ratio boundary is likely due to SDSS magnitude limitations.  We use a KS test to determine whether the overlapping portions of these distributions that would pass SDSS selection are identical.  Distributions have been normalized to 1.0 at peak, and only the portions of each distribution where the apparent magnitude corresponding to $L$ is above the SDSS detection threshold are shown.}
\label{fig:mvseddratio}
\end{figure}
Figure \ref{fig:mvseddratio} presents the distribution of quasar Eddington ratios at $0.2 < z < 0.4$ in the five most populous mass bins ($10^{7.75} < M/M_\odot < 10^{9.0}$).  The SDSS detection limit prevents lower-mass bins from containing objects at lower Eddington ratios than shown.  The distributions show a clear trend to higher $L/L_{Edd}$ at lower masses.  This trend is sufficiently strong that there is almost no overlap between bins of very disparate mass.  That these $L/L_{Edd}$ distributions are distinct can be shown via a comparison the pair of bins at $\Delta \log M$ greater than the $0.4$ dex mass uncertainty with the largest number of quasars in the overlapping region.  We use a KS test to quantify whether the overlapping portions of a pair of distributions are consistent with being drawn from the same distribution.  For the red (8.0-8.25 in $\log M/M_\odot$) and blue (8.5-8.75) distributions from Figure \ref{fig:mvseddratio}, the KS test yields a D value of 0.5862, and the probability that these are drawn from the same distribution is $3.6 \times 10^{-50}$, i.e., with $> 49\sigma$ confidence these are statistically different distributions.  Thus, the apparent SEB are not merely artifacts of small number statistics.  The difference in the medians of this pair of distributions separated by 0.5 dex in $M$ is 0.34 dex in $L/L_{Edd}$, or 0.16 dex in $L$.  The quasar luminosity distribution at $0.2 < z < 0.4$ is neither mass-independent nor linear in mass, but rather has some sub-linear dependence.  

The luminosity distributions in each mass bin appear to have similar shapes (Figure \ref{fig:mvslslice}).  To test this, the $L/L_{Edd}$ distributions are shifted to a common median value.  The KS test on the shifted red (8.0-8.25) and blue (8.5-8.75) distributions now yields a D value of just 0.0747, and the corresponding probability that these two distributions have the same shape is $0.358$.
\begin{figure}
  \epsfxsize=3in\epsfbox{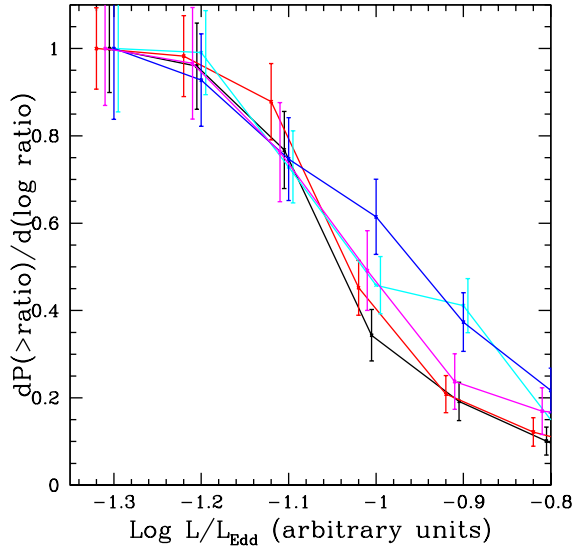}
\caption{The five distributions from Figure \ref{fig:mvseddratio} translated to match their median Eddington luminosites.  Distributions have been normalized to their peak.}
\label{fig:mvseddoverlap}
\end{figure}
Translating all five mass bins to a common median (Figure \ref{fig:mvseddoverlap}) confirms that these $L/L_{Edd}$ distributions are all similar.  This similarity is particularly striking because the total (sum at all masses) quasar luminosity function has a similar shape at different redshift \cite{Richards2006,Amarie2009}.  These similarities might hint at an underlying cosmic structure for quasar populations spanning a wide range of mass and redshift.  

We conclude that the SEB indeed sub-Eddington with very high statistical certainty.  Similarities in the shape of luminosity distributions at different black hole masses might provide a useful hint as to the origin of the SEB.

\subsection{SDSS selection}

SDSS does not generate spectroscopic data on every object in the survey, but rather only on those objects (including all quasar candidates) designated as worthy of followup based upon their photometry and observed apparent magnitudes in five spectral bands {\it ugriz} \cite{Richards2002}.  Every object identified as a quasar candidate is entered into the spectroscopic queue.  The DR5 SDSS spectroscopic footprint is a subset of the DR5 photometric footprint.  Regardless of the original photometric classification, any object identified as a quasar from its spectrum is included in the quasar catalogue, while any objects originally targeted as a quasar whose spectrum shows otherwise is excluded.  However, quasars miscategorized likely have no spectroscopic data, and therefore will be missing from the QSO catalogue.

The SDSS catalogue contains a large number of `serendipitous' quasars, selected often for an unusual color or FIRST match \cite{Richards2002}.  As a result, SDSS detection is only complete for objects brighter than 19.1 in $i$ band, but includes fainter objects.  The sample used in this paper includes all SDSS objects for which masses could be determined, including serendipitous objects.  However, because the SEB is determined by the most luminous and therefore brightest quasars at each redshift, the serendipitous sample has negligible effect upon the determination of the SEB.

As part of the original target selection study, Richards et al. (2002)\nocite{Richards2002} compared SDSS photometric quasar selection to known quasar catalogues, finding that 92.7\% of 2096 known quasars, including 94.5\% of the 1540 in their `bright' subsample, were correctly targeted as quasars by SDSS.  Simulations suggest that the true quasar completeness is closer to 90\%.  There is a gap in the SDSS catalogue around $2.5 < z < 3$, where the quasar locus crosses the stellar locus.  This gap does not affect our analysis of the SEB, which is restricted to $z < 2$.

The SEB occurs among the brightest quasars in each redshift bin, so saturation in SDSS might present a problem.  Since 3C273, at a redshift of just 0.16, is a 13th-magnitude quasar \cite{Schmidt1963}, we must ensure that this luminosity cutoff is not merely an artifact of saturation.  Jester et al. (2005)\nocite{Jester2005} compares SDSS with the Bright Quasar Survey (BQS)\cite{Schmidt1983} derived from the Palomar-Green survey.  Of the 51 objects in BQS that lie within the DR3 footprint, 29 are in the SDSS DR3 catalogue, 3 are above $i=15.0$ and therefore excluded, and the remaining 19 have been identified as QSO candidates in DR3 photometry but were still in the spectroscopic queue at the time of release.  These 19 have all been targeted for inclusion in the DR7 catalogue, and some are included in the DR5 catalogue used here.  Jester et al. (2005)\nocite{Jester2005} also show that, while individual quasars are variable over 20-year time-scales, statistically BQS and SDSS show no systematic photometric bias with respect to each other.

Could the SEB occur because quasars at higher luminosity have different properties which cause them to be missed by the BQS?  In related work, Amarie \& Steinhardt (2009)\nocite{Amarie2009} compared the SDSS catalogue with the Veron-Cetty/Veron (VCV) catalogue \cite{Veron2006}, which is a compilation of quasars from every available source.  VCV would be a poor choice for a full completeness study because there is no guarantee of either high observational quality or statistical completeness in any region of the sky.  However, as VCV includes quasars selected by all current methods, any population that might cross the SEB will show up.  Amarie \& Steinhardt (2009)\nocite{Amarie2009} conclude that the VCV and SDSS loci are well-matched, outside of the $z=2.5-3.0$ band affected by the decrease in selection efficiency.  Also, while VCV contains tens of quasars with $i < 16$ at low redshift, no substantial population of SEB-crossing quasars are found.  
We can therefore conclude that the SDSS quasar sample is representative of quasars as a whole, and find no evidence that these two new luminosity bounds are introduced by artificial SDSS selection.

\subsection{Bolometric luminosity errors}

The SEB takes the form of a paucity of quasars at high luminosity.  So, if the luminosity of the most luminous quasars in each redshift bin were underestimated, this could simulate such a boundary.  The luminosity calculation has three principal components \cite{Richards2006b,Shen2008}.  First, the apparent magnitude in five colour bands is calculated from the SDSS photometry as part of the standard SDSS pipeline.  Then, a K correction is used to account to convert the apparent magnitudes at different redshift to a magnitude in $i$ band if the quasar were at $z=2$.  Finally, Richards et al. (2006)\nocite{Richards2006b} calibrate a bolometric correction to $M_i(z=2)$.  Amarie et al. (2009)\nocite{Amarie2009} consider the components in determining the bolometric luminosity for a given quasar and whether such a bias might be introduced.  They confirm that SDSS bolometric luminosities have statistical uncertainties of only a few percent for bright quasars.  Systematic uncertainties in bolometric corrections may be far larger.  However, explaining the SEB would require two peculiar properties of such a correction: (1) a systematic bias towards underestimating luminosities at high mass; and (2) a systematic bias with redshift that exactly cancels the mass bias such that the lowest-mass quasars at every redshift can reach their Eddington luminosity, while no quasars are ever super-Eddington.  This combination is sufficiently improbable that we should consider systematic errors in bolometric luminosity estimation a very unlikely explanation for the SEB.

\subsection{Spectroscopic errors}

Virial mass estimates take the form
given by eq. (\ref{vmeform})for the full width half maximum (FWHM) and continuum luminosity ($L_{\lambda}$) of different spectral lines.  Uncertainties in these measurements become uncertainties in the corresponding mass estimate.  
If errors in the FWHM and $L_\lambda$ are the source of the SEB, they must be systematic errors that lower $L/L_{Edd}$ by leading to artificially high mass estimates, i.e., overestimates of the FWHM, $L_\lambda$, or both.

Continuum luminosities would seem easy to measure for SDSS spectra, with pixels of width 70 km/s \cite{York2000}.  In practice, continuum fitting is straightforward near the H$\beta$ and C{\small IV} lines, but when using Mg{\small II}, if the red end of the emission line is near the 9200~\AA~upper limit of the SDSS spectrograph (i.e., for $z \sim 2$), it may be difficult to discern the extent of the iron `bump' / Balmer continuum (cf. Wills et al. 1985\nocite{Wills1985}), which must be subtracted as part of the continuum fit.  However, even a factor of two error in the continuum fit would lead to just a 0.14 dex error in virial mass estimation.

Fitting for the FWHM of spectral lines can be more difficult.  There are often as few as 10-15 pixels in an SDSS Mg{\small II} line profile.  The FWHM in the SDSS pipeline relies on a fit to the second moment of the line \cite{Shen2008}.  Some of the pixels may have poor signal-to-noise ratios or may coincide with an absorption line on the blue side of the line and should be discarded. 

To assess these affects, we performed an independent fit for a subsample of 3167 SDSS quasars.  We fit the broad line shape as the sum of two Gaussian components: a broad component and a narrow line component \cite{Hao2005}.  This technique is subject to a different set of biases than that of Shen et al. (2008), most importantly the possibility that spectral lines might be skewed or non-Gaussian.  Comparing these two techniques with different biases and sources of error gives a measure of systematic uncertainty in emission line fitting.  We performed these fits on a subset of the H$\beta$ mass sample, where the SEB is indicated most strongly.  

Our fitting routine is based upon LINEBACKFIT, part of the IDLSPEC2D utilities package \cite{Burles2006}.  For the H$\beta$ line, we consider the spectrum between rest wavelengths 4400--5200 \AA~and fit a fifth-order polynomial continuum in addition to Gaussian [O{\small III}] line profiles at 4959 \AA~and 5007\AA~as well as the double-Gaussian H$\beta$ profile.  Virial mass estimation requires the best-fitting continuum at 5100 \AA, so mass estimates are insensitive to [O{\small III}] fitting.  For the H$\beta$ line, where the continuum can be well fit on both sides of the spectral line, these line fits were very robust with regard to individual pixels.  This rudimentary continuum fit was incapable of fitting the Fe{\small II} lines near Mg{\small II}, and therefore would have produced poor mass estimates for our Mg{\small II} sample.

Simulating an absorption line by setting the flux in one pixel to zero somewhere in the spectral line resulted in virial mass changes of $\Delta~M_{BH} < 0.1$ dex for even the narrowest H$\beta$ lines in our sample (where one erroneous pixel represents the greatest fraction of the total line).  This approach also appears to be only weakly sensitive to the separation between the broad and narrow components, and to the continuum fit in the region surrounding the line peak. 

\begin{figure}
  \epsfxsize=3in\epsfbox{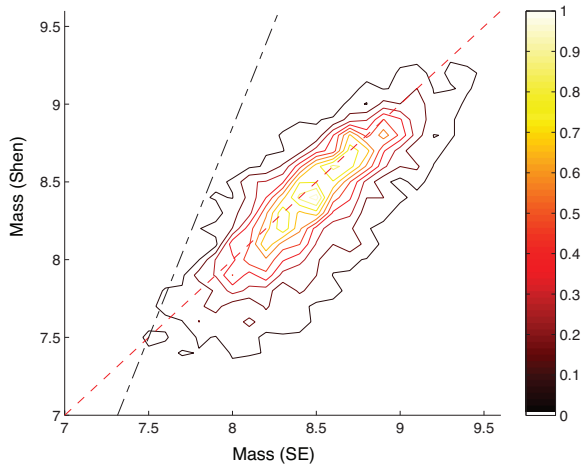}
\caption{A comparison of virial mass estimates from H$\beta$ lines using SDSS line fits (Shen et al. 2008) and Gaussian line fits from this paper.  The standard deviations is 0.30 dex.  The red dashed line is drawn where the two mass estimates are in agreement.  The black dashed line is drawn at the systematic mass offset required to move the SEB back to $L_{Edd}$ at $0.2 < z < 0.4$.  The number density has been normalized to the peak number density.}  
\label{fig:shencomp}
\end{figure}

In Figure \ref{fig:shencomp}, we show the correlation between our mass estimates ($M_{\textrm{SE}}$) and those of Shen et al. (2008; $M_{\textrm{Shen}}$).  There is a slight offset between these two mass estimates.  However, these two virial mass estimates based upon different line fitting techniques are still well-correlated, with a standard deviation of 0.30 dex in $M_{\textrm{Shen}}/M_{\textrm{SE}}$ for H$\beta$.  We conclude that the methods used to fit the line parameters entered into virial mass estimates do not substantially increase virial mass estimate uncertainty beyond the claimed $\sim 0.4$ dex.  Moreover, as shown in Figure \ref{fig:fitevolution}, the H$\beta$ and Mg{\small II} views of the SEB at $0.6 < z < 0.8$ are in strong agreement. 

The $10^9 M_\odot$ quasars at $0.2 < z < 0.4$ have a maximum luminosity of $L/L_E \sim 0.1$ (Figure \ref{fig:hb1}).  In order for the SEB to be spurious, these measurement errors would need to be at least $1$ dex at high mass, as well as have a systematic bias towards overestimating the mass.  The black dashed line in Figure \ref{fig:shencomp} is drawn with the relation that would be required to move the SEB back to the Eddington luminosity (Figure \ref{fig:hb1}).  This relation is clearly larger than the differences induced by these two different fitting methods.  We therefore find no evidence that measurement uncertainties in the virial mass estimates are introducing non-physical limits on the SDSS quasar locus.

\subsection{Virial mass estimates}
\label{subsec:vme}
If the virial mass estimates are incorrect then the quasar locus in the $M-L$ plane would be shifted.  However, in order for errors in virial mass estimates to explain the SEB requires that the masses are incorrect in such a way as to produce Figure \ref{fig:hb1} by skewing the locus such that quasars at high mass reach either their Eddington luminosity or the SDSS bright-object cutoff.  

In particular, the most luminous quasars at $0.2 < z < 0.4$ have $L \sim 10^{46.1}$erg/s (Figure \ref{fig:hb1}).  In order to correspond to the Eddington luminosity, no quasars in this redshift range could have a black hole mass larger than $10^{8.0}M_\odot$.  This would imply that a majority of quasars have vastly overestimated masses, including many central black holes with masses overestimated by over a factor of 10 and one by at least a factor of 100.  Similarly (Figure \ref{fig:allzcontour}), at a redshift of $1.0 < z < 1.2$, the most luminous quasars have $L \sim 10^{47.2}$ erg/s, implying $M_{BH} \leq 10^{9.1}M_\odot$.  This would correspond to typical overestimates of at least 0.5 dex and for some objects 0.8 dex.  

A comparison of virial mass estimates to reverberation mapping-based masses shows a typical uncertainty closer to 0.4 dex \cite{Vestergaard2006}.  The virial mass estimates must then have additional imprecision.  Vestergaard \& Peterson\nocite{Vestergaard2006} (2006) suggest these are primarily statistical in nature.  If this is true, then if many masses in Figure 3 are overestimated by a factor of ten, we should also see masses underestimated by a factor of ten and therefore objects that appear to be at $L = 10L_E$.  Since we do not, using virial mass mis-estimation as an explanation for these new boundaries would require a systematic component as well.  We attempt to evaluate both statistical and systematic uncertainties in virial mass estimates below.

\subsubsection{Virial mass estimates: statistics}

We can make an independent estimate of the statistical error in $M_{BH}$ by considering the low-mass end of the $M-L$ plane at each redshift, where some quasars reach their Eddington luminosity.  Statistical uncertainties should result in some quasars lying at a lower mass than is physically allowed and thus appearing to be above $L_{Edd}$.  We can estimate the true statistical uncertainty in $M_{BH}$ from the width of this falloff in quasar number density.

\begin{figure}
  \epsfxsize=3in\epsfbox{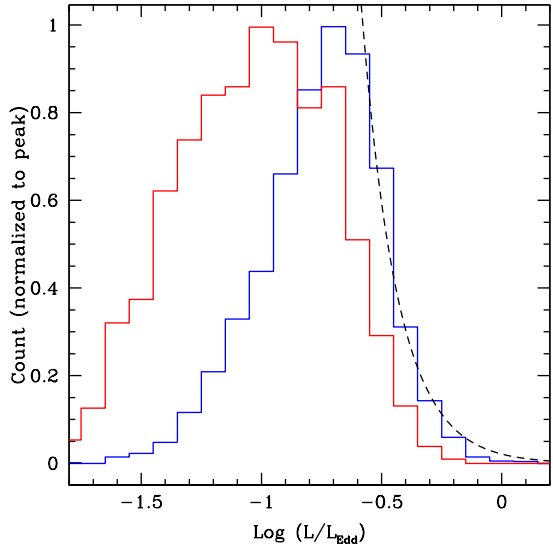}
\caption{Quasar number density using H$\beta$ masses at $0.6 < z < 0.8$ (red) and Mg{\small II} masses at $1.0 < z < 1.2$ (blue) as a function of Eddington ratio for quasars within one magnitude of the SDSS detection limit.  The dashed line is an exponential decay with an e-folding of $0.15$ dex in $L/L_E$.}
\label{fig:densityturnoff}
\end{figure}

The quasar populations within one $i$-band magnitude of the SDSS detection limit (Figure \ref{fig:densityturnoff}) do not reach $L_{Edd}$, but decline sharply at high Eddington ratio, with an e-folding number density rate of $\sim 0.15$ dex in $L/L_{Edd}$.  These quasars approach $L_{Edd}$ more closely than at higher luminosity, where quasars are bounded by the SEB.  This $\sim 0.15$ dex falloff is a combination of statistical uncertainty (which should be Gaussian), possible systematic effects, and possibly a real, underlying decline in the quasar $L/L_{Edd}$ distribution associated with the Eddington limit.  Therefore, without knowing the details of underlying physical cause of the decline, there is a $\sim 0.15$ dex  upper limit for the statistical uncertainty in the Shen et al. (2008) virial mass estimates.  If the underlying physical cause is a gradual rather than a sharp decline in number density, the maximum possible statistical uncertainty in virial mass estimates will be reduced.

$0.15$ dex is notably smaller than the uncertainty estimated by Vestergaard \& Peterson (2006)\nocite{Vestergaard2006} in a comparison of reverberation and H$\beta$-based virial masses for the same set of objects.  This suggests that the disagreements between the reverberation mapping and virial mass estimates have a substantial systematic component.  A closer examination of the Vestergaard \& Peterson (2006)\nocite{Vestergaard2006} mass estimate comparison suggests that reverberation mapping-based estimates at high mass might be larger than virial mass estimates.  This would result in a stronger SEB (i.e., one with a lower slope) than reported in Table {\ref{table:fitparams}}.

The statistical uncertainty in $M_{BH}$ can also be estimated by comparing different virial mass scaling laws at redshifts where two lines can be used.  At $0.6 < z < 0.8$, there are 3505 quasars in common between the H$\beta$ and Mg{\small II} samples.  Since the SEB occurs at the high-luminosity end of the quasar sample, if is due to statistical uncertainties in mass estimation there should be a higher dispersion between different mass estimates for more luminous quasars than for the population as a whole.

At $0.6 < z < 0.8$, the mass dispersion for all quasars common to the H$\beta$ and Mg{\small II} samples is 0.20 dex, while the mass dispersion for the 10\% most luminous quasars is 0.16 dex.   These dispersions are substantially smaller than the $\sim 1.0$ dex required to remove the SEB.  The smaller dispersion for luminous quasars likely occurs because brighter objects are typically accompanied by higher signal-to-noise spectra.  Not only are individual H$\beta$ and Mg{\small II}-based mass estimates well-correlated for bright quasars, but so is the resulting SEB.  At $0.6 < z < 0.8$, the best-fitting SEB shown in Table \ref{table:fitparams} has slope $\alpha = 0.60 \pm 0.06$ for H$\beta$ masses and $\alpha = 0.61 \pm 0.10$ for Mg{\small II} masses.  

Statistical uncertainty could produce high-mass outliers that might contribute to high-mass deviations from the best-fitting line, but a correction of the SEB to $L_{Edd}$ would require an overestimate of at least 1.0 dex for many quasars.  These estimates for the statistical uncertainty in $M_{BH}$ range from just 0.15--0.4 dex, too small to artificially produce the SEB.  Further, statistical uncertainty would broaden the locus, not change the slope, and the SEB has a slope at least $3.7\sigma$ away from the $\alpha = 1$ of $L_{Edd}$ in every redshift bin.  
We conclude that while statistical uncertainty is a factor in the final shape of the quasar locus at the 0.15 dex level, we find no evidence from any of our tests that statistical uncertainty could be responsible for the $\sim 1$ dex SEB shown in Figure \ref{fig:hb1}.

\subsubsection{Virial mass estimates: systematics}

The falloff in Figure \ref{fig:densityturnoff} starts at $\log L/L_{Edd} \sim -0.5$ at low mass and $\log L/L_{Edd} \sim -1.0$ at high mass for H$\beta$.  If some quasars reach their Eddington luminosities, this early falloff suggests that masses could be systematically overestimated.  This might mean masses are overestimated by as much as $0.5-1.0$ dex.  A systematic overestimate in virial mass estimates would likely also require a systematic overestimate in the reverberation mapping-based estimates against which they are calibrated, although a sufficiently small overestimate might lie within the 0.4 dex uncertainties between the two methods.  It appears that systematic errors of of $\sim 1$ dex in virial mass estimation cannot be entirely ruled out with this analysis.  

However, the SEB slope requires not just a shift in $M_{BH}$ but a systematic, mass-dependent change in the mass estimate.  Low-mass quasars cannot have mass overestimates large enough for their luminosities to exceed $L_{Edd}$, so that a correction of the SEB might require a 1.5 dex overestimate at the high-mass end with only a 0.5 dex overestimate at low mass.  Since luminosities at $0.2 < z < 0.4$ (Figure \ref{fig:hb1}) run over a range of only 1.3 dex, the actual underlying black hole masses must run over a range no larger than 1.3 dex themselves, or else the SEB cannot really have slope $\alpha=1$.  If true, this would replace the SEB with a different surprising feature of the quasar $M-L$ distribution.

For Mg{\small II} masses in particular, Onken \& Kollmeier (2008)\nocite{Onken2008} examine SMBH for which both H$\beta$ and Mg{\small II} masses are available and, assuming the H$\beta$ mass is the better indicator, find that the Mg{\small II}-based $M_{BH}$ may be overestimated at high Eddington ratio and underestimated at low Eddington ratio.  Correcting for this effect will drive low-mass objects at high Eddington ratios to even lower masses and therefore closer to $L_{Edd}$, but high mass objects at low Eddington ratios to higher masses and lower Eddington ratios.  Therefore, such a correction could produce a stronger, more sub-Eddington boundary. 

Risaliti, Young, \& Elvis (2009)\nocite{Risaliti2009} also compare H$\beta$ and Mg{\small II} masses, finding that 
\begin{equation}
\log[M_{BH}(H\beta)]=1.8\times\log[M_{BH}(Mg{\small II})]-6.8.
\end{equation}
This correction acts in the same direction as the correction proposed by Onken \& Kollmeieier and will change the SEB slope $\alpha$ in each Mg{\small II} mass bin from $L = \alpha M$ to $L = \alpha^\prime (1.8 M^\prime)$, for a reduction in $\alpha$ by a factor of 1.8 (Figure \ref{fig:fitevolution}).
This correction produces a 3$\sigma$ disagreement between the H$\beta$ and Mg{\small II} SEB slopes at $0.6 < z < 0.8$, although since these two samples comprise a very similar set of objects, it is almost certainly incorrect to treat their SEB uncertainties as statistical in nature.  This result seems to require that the brightest quasars at each mass need smaller Mg{\small II} mass corrections than fainter ones.  Perhaps high signal-to-noise spectra yield more reliable Mg{\small II} masses.  We conclude that specific, known systematic errors in virial mass estimation might take the SEB slopes closest to Eddington and move them further sub-Eddington.  Any new correction that might bring SEB slopes to $L_{Edd}$ would replace the SEB with a different surprising feature of the quasar distribution in the $M-L$ plane.

\subsection{Underlying mass distribution}
\begin{figure}
  \epsfxsize=3in\epsfbox{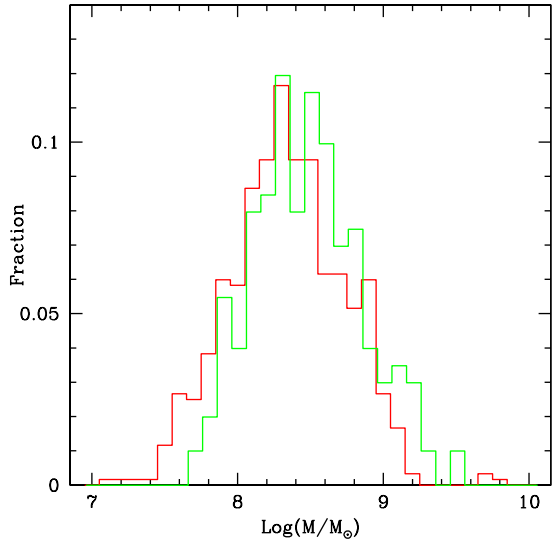}
\caption{The quasar mass distribution for two different luminosity bins in the $0.2<z<0.4$ sample shown in Figure {\ref{fig:hb1}}.  We divide the sample by luminosity into bins with $10^{45.3} < L/L_\odot < 10^{45.5}$ (601 objects, red) and $10^{45.5} < L/L_\odot < 10^{45.7}$ (203 objects, green).  The higher-luminosity objects are on average 0.20 dex higher in luminosity but only 0.09 dex higher in mass.}
\label{fig:mfl}
\end{figure}

One may wonder whether we have simply re-established the well-known decline of the luminosity function of quasars at high luminosity \cite{Schmidt1968,Richards2006}, and whether the `missing' objects simply reflect the underlying mass distribution.  In Figure \ref{fig:mfl} we compare the mass distribution of black holes in two well-populated luminosity bins at the same redshift ($0.2 < z < 0.4$) separated by 0.2 dex.  
If the SEB were the expression of mass turnoff, objects in the bin 0.2 dex higher in luminosity would also be 0.2 dex more massive, just that there would be fewer at higher mass and luminosity.  In fact, the higher-luminosity objects are on average just 0.09 dex more massive.  Further, a KS test comparing the mass distributions with a shift of 0.09 dex results in a probability of 47.2\% that they are identical, while a shift of 0.2 dex results in a probability of only 6.0\% that the two are drawn from the same distribution.  Not only does the SEB have a slope $\alpha < 1$, but for the entire quasar population, an increase in mass corresponds to a sub-linear increase in luminosity.

In particular, the SEB is not simply the expression of mass turnoff, as the luminosity increase is only slightly correlated with mass increase at the bright end.  This does raise the possibility, discussed in Paper II \cite{Steinhardt2009b}, that we can probe quasar turnoff using SDSS data. 
 
The SEB is an expression of the relative scarcity of high-mass, high-Eddington ratio objects compared to high-mass, low-Eddington ratio and low-mass, high-Eddington ratio objects at each redshift.  While the mass estimate does depend upon luminosity (as well as the FWHM of a broad emission line), an increase of 1 dex in luminosity leads directly to an increase of just 0.5 dex in mass (Eq. \ref{vmeform}), not the $\sim 1.7$ dex increase required for a typical SEB slope.  Since the SEB is determined using objects only at a common redshift, there are no changes in the comoving volume to account for in evaluating this scarcity.  

Because both masses and (bolometric) luminosities are derived quantities, the SEB could be alternatively interpreted as either (1) a physical clue about the components of these derived quantities (e.g., at high masses, an increase in luminosity might in every case be accompanied by a much larger than expected increase in the FWHM of broad emission lines), (2) a failure in deriving these quantities (e.g., virial mass estimation breaks down on the high-mass end at each redshift in exactly such a manner as to produce this skew but is correct at higher redshift when identical masses lie on the low-mass end), or (3) a true lack of high-mass, high-luminosity quasars.  We have argued that (3) should be the favored interpretation because of the fine-tuning required if (1) or (2) is to apply only the highest-mass objects at each redshift.  However, all three interpretations would require that something physical is different about high-mass, high-luminosity quasars in each cosmological epoch and therefore any interpretation presents a theoretical puzzle.

We have considered six potential explanations for these new limits without introducing new quasar physics, as summarized in Table \ref{table:tests}.  
\begin{table}
\caption{Possible explanations for the sub-Eddington boundary considered in \S~{\ref{sec:tests}}.}
\begin{tabular}{|c|c|}
\hline 
Explanation & Plausible? \\
\hline 
Statistical Insignificance & No \\
SDSS Selection & No \\
Luminosity Errors & No \\
Line Measurement Errors & No \\
Virial Masses: Statistics & No (maybe at high $M$) \\
Virial Masses: Systematics & Unlikely \\
Quasar Turnoff & No \\
\hline 
\end{tabular}
\label{table:tests} 
\end{table}
A large, unknown systematic error in virial $M_BH$ measurements is the most plausible explanation by comparison for how the SEB might be produced without additional underlying physics, but requires that virial mass estimates are $\sim 1$ dex incorrect, reverberation mapping is $\sim 1$ dex incorrect, and that quasars at low redshift lie within a spread of just 1.3 decades of central black hole mass.  It would appear likely that the explanation for these new limits lies in new theory rather than observational errors.

\section{Discussion}
\label{sec:discussion}

Quasar catalogues such as the Sloan Digital Sky Survey are now large enough to yield useful statistical conclusions after being subdivided.  In this work, we have explored just a few of the many ways in which the catalogue might be divided.  It is immediately apparent that when the quasar distribution is considered at any given redshift, the most massive central black holes do not accrete at their Eddington luminosity, but rather all fall well short of $L_E$.  While there are many measurements that contribute to these quasar distributions, each with its own set of assumptions, a close examination reveals no evidence that any these are responsible for massive quasars remaining sub-Eddington.  Rather, it appears that this `sub-Eddington boundary' is a new physical limit for quasars.  Thus, quasar accretion must be more complicated than had been previously thought.

The sub-Eddington boundary helps to recast the problem in a two-dimensional form.  This form emphasizes the luminosity as being a function of the mass and accretion rate, fundamental properties of the quasar, in a non-trivial way.  Quasar luminosity functions and quasar mass functions are both one-dimensional projections of the mass-luminosity plane.  The full two-dimensional quasar distribution contains complexities difficult to discern from either of these projections.

The sub-Eddington boundary presents a theoretical puzzle.  Not only must an improved theory explain the sub-Eddington boundary, but it must also explain its detailed shape and explain the evolution of that shape with redshift.  While the slope of the sub-Eddington boundary may not vary greatly with redshift, the mass scale at which the boundary becomes relevant is larger at higher redshift.  The results in this paper can be used to develop a series of tests which should be capable of discriminating between models.

Given the complications introduced by the sub-Eddington boundary, the case could be made that we really understand surprisingly little about the supermassive black holes that appear to be at the center of nearly every galaxy.  Not only is their seeding mechanism unknown, but as shown in this paper, the growth mechanism is quite poorly understood, contrary to expectations.  The Eddington limit is relevant at low black hole masses, but is only part of the story.  The sub-Eddington boundary developed in this work is the latest addition to a growing collection of puzzles regarding every phase of the evolution of galactic nuclei and surrounding regions. 

The authors would like to thank Mihail Amarie, Forrest Collman, Doug Finkbeiner, Margaret Geller, Lars Hernquist, Gillian Knapp, Avi Loeb, Ramesh Narayan, Jerry Ostriker, and Michael Strauss for valuable comments.  This work was supported in part by Chandra grant number GO7-8136A (Chandra X-ray Center).


\begin{thebibliography}{99xx}

\bibitem[Abazajian et al. 2005]{SDSSDR3}Abazajian K., Adelman J., Agueros M., et al., 2005, AJ, 129, 1755

\bibitem[Amarie et al. 2009]{Amarie2009}Amarie, M., Steinhardt, C. L., 2009, in preparation

\bibitem[Begelman 2002]{Begelman}Begelman M.~C., 2002, Astrophysical Journal Letters, 568, L97 

\bibitem[Bentz et al. 2009]{Bentz2009}Bentz M., Peterson B.~M., Netzer H., Pogge R. W., Vestergaard M., 2009, Astrophysical Journal, submitted; preprint astro-ph/0812.2283

\bibitem[Burles \& Schlegel 2006]{Burles2006}Burles S., Schlegel D., 2006, in preparation

\bibitem[Ferrarese \& Merritt 2000]{msigma1}Ferrarese L., Merritt D., 2000, ApJ, 539L, 9

\bibitem[Ferrarese \& Ford 2005]{Ferrarese2005}Ferrarese L., Ford H., 2005, Space Science Rev., 116, 523

\bibitem[Gavignaud et al. 2008]{Gavignaud2008}Gavignaud I., Wisotzki L., Bongiorno A. et al., 2008, Astronomy \& Astrophysics, accepted; preprint astro-ph/0810.2172

\bibitem[Gebhardt et al. 2000]{msigma2}Gebhardt K., Kormendy J, Ho L. et al., 2000, ApJ, 539L, 13

\bibitem[Goldschmidt et al. 1999]{Goldschmidt1999}Goldschmidt, P. Kukula, M. J., Miller, L., Dunlop, J. S., 1999, ApJ, 511, 612

\bibitem[Hao et al. 2005]{Hao2005}Hao L., Strauss M. A., Tremonti C. A.\ et al., 2005, AJ, 129, 1783 

\bibitem[Jester et al. 2005]{Jester2005}Jester S., Schneider D. P., Richards G. T.\ et al., 2005, AJ, 130, 873

\bibitem[Jiang et al. 2006]{Jiang2006}Jiang L., Fan X., Ivezic Z., Richards G. T., Schneider D. P., Strauss M. A., Kelly B. C., 2007, AJ, 656, 680

\bibitem[Juneau et al. 2005]{Juneau}Juneau S., Glazebrook K., Crampton D.\ et al., 2005, Astrophysical Journal Letters, 619, L135 

\bibitem[Kollmeier et al. 2006]{Kollmeier2005}Kollmeier J., Onken C. A., Kochanek C. S.\ et al., 2006, ApJ, 648, 128

\bibitem[Marconi et al. 2009]{Marconi2009}Marconi A., Axon D., Maiolino R., Nagao T., Pietrini P., Robinson A., Torricelli G., 2009, Astrophysical Journal, submitted; preprint astro-ph/0809.0390

\bibitem[McLure \& Jarvis 2002]{McLure2002}McLure R.J., Jarvis, M.J., 2002, MNRAS, 337, 109

\bibitem[McLure \& Dunlop 2004]{McLure2004}McLure R.J., Dunlop, J.S., 2004, MNRAS, 352, 1390

\bibitem[Merritt \& Ferrarese 2001]{Merritt2001}Merritt D., Ferrarese, L., 2001, MNRAS, 320, L30
 
\bibitem[Miller, Rawlings, \& Saunders 1993]{Miller1993}Miller P., Rawlings S., Saunders R., 1993, MNRAS, 263, 425

\bibitem[Onken \& Kollmeier 2008]{Onken2008}Onken C. A., Kollmeier J. A., 2008, Astrophysical Journal Letters, submitted; preprint astro-ph/0810.1950

\bibitem[Peterson 2008]{Petersonbook}Peterson B., 2008, An Introduction to Active Galactic Nuclei (Cambridge University Press: Cambridge)

\bibitem[Peterson \& Horne 2004]{Peterson2004}Peterson, B. M., Horne K., 2005, in Planets to Cosmology: Essential Science in Hubble's Final Years, M. Livio ed.

\bibitem[Richards et al. 2002]{Richards2002}Richards G.T., Fan X., Newberg H. et al., 2002, AJ, 123, 2945

\bibitem[Richards et al. 2006a]{Richards2006}Richards G. T., Strauss M. A., Fan X. et al., 2006, AJ, 131, 2766

\bibitem[Richards et al. 2006b]{Richards2006b}Richards, G. T., Lacy M., Storrie-Lombardi, L. J. et al., 2006, ApJ Supp., 166, 470

\bibitem[Risaliti, Young, \& Elvis 2009]{Risaliti2009}Risaliti G., Young M., Elvis M., 2009, accepted to Astrophysical Journal Letters; preprint astro-ph/0906/1983

\bibitem[Schmidt 1963]{Schmidt1963}Schmidt M., 1963, Nature, 197, 1040

\bibitem[Schmidt 1968]{Schmidt1968}Schmidt M., 1968, AJ, 151, 393

\bibitem[Schmidt \& Green 1983]{Schmidt1983}Schmidt M., Green R., 1983, ApJ, 269, 352

\bibitem[Schneider et al. 2007]{SDSSDR5}Schneider D. P., Hall P. B., Richards G. T. et. al., 2007, AJ, 134

\bibitem[Shapiro \& Teukolsky 1983]{Shapiro}Shapiro, S. L., \& Teukolsky,
S. A., 1983, Black Holes, White Dwarfs, and Neutron Stars: The Physics of
Compact Objects (Wiley: New York), p. 396

\bibitem[Shen et al. 2008]{Shen2008}Shen Y., Greene J. E., Strauss M. A., Richards G. T., Schneider D. P., 2008, ApJ, 680, 169

\bibitem[Smith et al. 2005]{2dF}Smith R.~J., Croom S.~M., Boyle B.~J., Shanks T., Miller L., Loaring N.~S., 2005, MNRAS, 359, 57

\bibitem[Spergel et al. 2006]{WMAP3}Spergel D. N., Bean R., Dore O. et al, 2006, Astrophysical Journal Supplement, 170, 377

\bibitem[Springel et al. 2005]{Springel}Springel V., Di Matteo T., Hernquist L., 2005, MNRAS, 361, 776 

\bibitem[Steinhardt \& Elvis 2009]{Steinhardt2009b}Steinhardt, C. L. \& Elvis, M., 2009, in preparation

\bibitem[Thacker et al. 2006]{turnoffreview}Thacker R. J, Scannapieco E., Couchman, H. M. P., 2006, ApJ, 653, 86

\bibitem[Trump et al. 2009]{Trump2009}Trump, J. R., Impey, C. D., Kelly, B. C. et al., 2009, ApJ, 700, 49

\bibitem[Veron-Cetty \& Veron 2006]{Veron2006}Veron-Cetty M.-P., Veron P., 2006, CDS/ADC Coll. Elec. Cat., 7248, 0

\bibitem[Vestergaard \& Peterson 2006]{Vestergaard2006}Vestergaard, M., Peterson, B., 2006, ApJ, 641, 689

\bibitem[Wampler \& Ponz 1985]{Wampler1985}Wampler E.J., Ponz D., 1985, ApJ, 298, 448

\bibitem[Willott, McLure \& Jarvis 2003]{Willott2003}Willott, C. J., McLure R. J., Jarvis M. J., 2003, Astrophysical Journal Letters, 587, L15

\bibitem[Wills, Netzer \& Wills 1985]{Wills1985}Wills B. J., Netzer H., Wills D., 1985, ApJ, 288, 94

\bibitem[York et al. 2000]{York2000}York D. G., Adelman J., Anderson J. E. et al., 2000, AJ, 120, 1579

\bibitem[Yu \& Tremaine 2002]{Yu2002}Yu Q., Tremaine S., 2002, MNRAS, 335, 965

\end{thebibliography}
\end{document}